# Nonequilibrium Casimir pressures in liquids under shear

J.M. Ortiz de Zárate[1], T.R. Kirkpatrick[2], and J.V. Sengers[2,a]

[1]Departamento de Estructura de la Materia, Facultad de Física, Universidad Complutense, 28040 Madrid, Spain
[2]Institute for Physical Science and Technology, University of Maryland, College Park, Maryland 20742, USA

[a]e-mail : sengers@umd.edu

**Abstract**. In stationary nonequilibrium states coupling between hydrodynamic modes causes thermal fluctuations to become long ranged inducing nonequilibrium Casimir pressures. Here we consider nonequilibrium Casimir pressures induced in liquids by a velocity gradient. Specifically, we have obtained explicit expressions for the magnitude of the shear-induced pressure enhancements in a liquid layer between two horizontal plates that complete and correct results previously presented in the literature. In contrast to nonequilibrium Casimir pressures induced by a temperature or concentration gradient, we find that in shear nonequilibrium contributions from short-range fluctuations are no longer negligible. In addition, it is noted that currently available computer simulations of model fluids in shear observe effects from molecular correlations at nanoscales that have a different physical origin and do not probe shear-induced pressures resulting from coupling of long-wavelength hydrodynamic modes. Even more importantly, we find that in actual experimental conditions, shear-induced pressure enhancements are caused by viscous heating and not by thermal velocity fluctuations. Hence, isothermal computer simulations are irrelevant for the interpretation of experimental shear-induced pressure enhancements.

## 1 Introduction

When large and long-range fluctuations are present, they will induce forces in confined fluids [1]. They are commonly referred to as Casimir-like forces in analogy to forces induced by vacuum fluctuations between two conducting plates [2]. Well-known examples are Casimir forces due to critical fluctuations [3] or due to long-range correlations in condensed systems with Goldstone modes [1, 4]. It has now been well established that even longer-range thermal fluctuations exist in fluids in nonequilibrium states [5]. The physical reason is that the presence of a gradient breaks the symmetry and causes a coupling between long-wavelength hydrodynamic modes, which are especially important in the convective nonlinear terms in the Navier Stokes equations [6].

In this paper we consider Casimir forces due to long-range thermal velocity fluctuations in laminar fluid flow [7-9]. For the case of a liquid layer subjected to a stationary velocity gradient between two parallel plates, we have obtained explicit expressions for the shear-induced pressure enhancements which correct and extend results obtained by previous investigators [10-13]. We provide quantitative estimates for the magnitude of these shear-induced Casimir pressures. In addition, we present an extended kinetic theory approach to compare nonequilibrium Casimir pressures induced by long-range thermal fluctuations with nonequilibrium pressures resulting from short-range thermal fluctuations. We clarify an essential difference between the Casimir pressures caused by macroscopic long-range fluctuations and pressures resulting from fluctuations at nanoscales which are observed in computer simulations. Finally, we shall point out that in actual experimental conditions, observed shear-induced pressure enhancements are caused by viscous heating, and not by thermal velocity fluctuations.

We shall proceed as follows. Continuing an approach adopted in some previous publications to determine the intensity of thermal velocity fluctuations in laminar liquid flow [14-17], we start in Section 2 from a fluctuating Orr-Sommerfeld equation for the wall-normal velocity fluctuations and from a fluctuating Squire equation for the wall-normal vorticity fluctuations. The solutions of these equations are then converted into expressions for the fluctuations of all velocity components, not only in the wall-normal direction, but also in the stream-wise and the span-wise directions. The procedure for solving these fluctuating hydrodynamics equations is also indicated in Section 2, but the mathematical details are presented in Appendices. We have solved the fluctuating hydrodynamics equations both in the absence and in the presence of boundary conditions. The solutions in the absence of boundary conditions are obtained from previous publications [14, 15], but for an evaluation of shear-induced Casimir forces in confined liquid layers it is essential to include finite-size effects. In previous publications we have considered liquid layers confined between two rigid surfaces where no slip occurs. However, in the case of such rigid boundaries it is very difficult to get an exact solution [16, 17] and in practice we have previously settled for an approximate solution in a so-called Galerkin approximation [14, 15]. In the present paper we have adopted periodic boundary conditions for two reasons. First, for periodic boundary conditions we are able to get an exact solution as was possible for the case without boundary conditions. Second, periodic boundary conditions are commonly adopted in computer simulations [18-25].

In Section 3 we present the elements of the nonequilibrium pressure tensor thus obtained from the fluctuating hydrodynamics equations. We find a scaling relation for the shear-induced pressure enhancement in terms of a function that, for a given set of boundary conditions, only depends on the Reynolds number Re. We present exact results for the shear-induced pressure enhancements both in the limit of large and of small Re and also discuss the nature of the crossover from small Re to large Re behavior. Specifically, we find that for laminar-flow conditions, finite-size effects always need to be included.

In Section 4 we discuss the magnitude of the shear-induced pressure enhancements and, in particular, show how our new results correct and extend results previously obtained by some other investigators [10-13]. We also present in Section 4 estimates of the shear-induced pressure enhancements for realistic experimental conditions. It turns out that, in contrast to nonequilibrium Casimir pressures induced by a temperature gradient [26], for nonequilibrium Casimir pressures induced by a velocity gradient contributions from short-ranged velocity fluctuations cannot be neglected. In Section 5 we review the currently available computer simulations for determining shear-induced pressure enhancements. A problem is that molecular dynamics simulations observe correlations at nanoscales which have a different physical origin than the nonequilibrium pressures arising from the long-range velocity fluctuations.

In Section 6 we provide estimates of pressure enhancements from possible viscous heating effects in real experimental conditions. We find that in real experiments these viscous effects will be dominant.

Our principal conclusions are summarized in Section 7.

## 2 Fluctuating hydrodynamics in laminar fluid flow

To elucidate the role of nonequilibrium velocity fluctuations in laminar flow, we consider the simplest case, namely that of a liquid under isothermal incompressible laminar flow (thus with uniform temperature $T$ and density $\rho$) between two horizontal boundaries, commonly referred to as planar Couette flow. To maintain consistency with our previous analysis of nonequilibrium velocity fluctuations [14-17], we continue using here the nomenclature of Drazin and Reid [27], sometimes referred as the meteorological convention [28], as indicated schematically in Fig. 1. Specifically, we use a coordinate system where the $x$ coordinate is in the stream-wise direction, the $y$ coordinate in the span-wise direction, and the $z$ coordinate in the wall-normal direction. The liquid layer is confined between two horizontal boundaries located at $z = \pm L$ moving with constant velocities $\pm U$ in the $x$ direction. The local fluid velocity can be decomposed as $\mathbf{v} = \mathbf{v}_0(z) + \delta\mathbf{v}$, where $\mathbf{v}_0 = \{\gamma z, 0{,}0\}$ is the average velocity depending on the shear rate $\gamma = U/L$ with a component only in the stream-wise direction $x$, and where $\delta\mathbf{v}(\mathbf{r}, t)$ is a fluctuating-velocity contribution dependent on the location $\mathbf{r}(x, y, z)$ and on the time $t$. As is common in the statistical-physics literature on the subject [12], we assume isothermal fluid flow and neglect here any viscous-heating effects, but they will be considered later in Section 6.

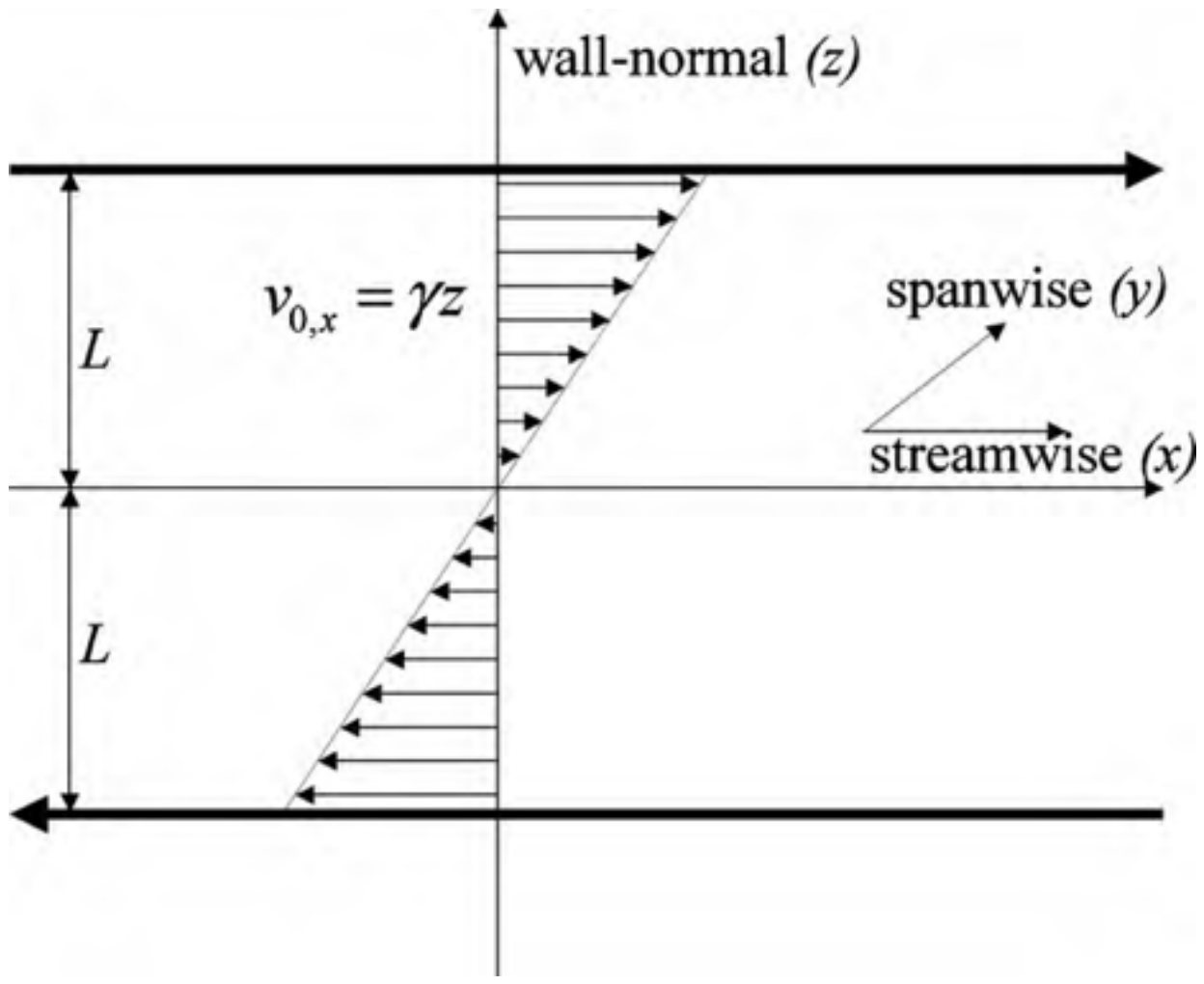

**Fig. 1.** Schematic representation of planar Couette flow.

Our task is to evaluate the nonequilibrium contribution $\delta\mathrm{P}(\mathbf{r})$ to the pressure tensor arising from the long-range nonequilibrium velocity fluctuations:

$$\delta\mathrm{P}(\mathbf{r}) = \rho\langle\delta\mathbf{v}(\mathbf{r})\delta\mathbf{v}(\mathbf{r})\rangle_{\mathrm{NE}}, \tag{1}$$

where the average is taken over the stationary nonequilibrium state which is independent of the time $t$. In principle there are also contributions from nonequilibrium density and internal-energy contributions, but the dominant contribution to the nonequilibrium pressure tensor arises from the velocity fluctuations [10]. The diagonal elements $\delta p_{ii} = \rho\langle\delta v_i \delta v_i\rangle$ contribute to the shear-induced pressure enhancement, such that $\delta p = \frac{1}{3}\left(\delta p_{xx} + \delta p_{yy} + \delta p_{zz}\right)$. For reasons of symmetry, the off-diagonal elements all vanish except for $\delta p_{xz} = \rho\langle\delta v_x \delta v_z\rangle$, yielding a fluctuation-induced contribution to the shear viscosity $\eta$ [12].

The relevant linearized fluctuating hydrodynamics equations for the fluctuations $\delta\mathbf{v}(\mathbf{r},t)$ and $\delta p(\mathbf{r},t)$ of the velocity and the pressure at location **r** and time $t$ around their mean values $\mathbf{v}_0 = \{\gamma z, 0,0\}$ and $p = p_0$ are [6, 29]

$$\nabla\cdot\delta\mathbf{v} = 0, \tag{2}$$

$$\frac{\partial(\delta\mathbf{v})}{\partial t} + \gamma z\frac{\partial(\delta\mathbf{v})}{\partial x} + \delta v_x\,\gamma\hat{\mathbf{x}} = -\frac{1}{\rho}\boldsymbol{\nabla}\delta p + \nu\nabla^2\delta\mathrm{v} + \frac{1}{\rho}\boldsymbol{\nabla}\cdot\delta\Pi, \tag{3}$$

where $\nu$ is the kinematic viscosity and where $\delta\Pi(\mathbf{r},t)$ is a random fluctuating tensor whose autocorrelation function is given by a fluctuation-dissipation theorem, which for incompressible (divergence-free) flow reads [6]:

$$\langle\delta\Pi_{ij}(\mathbf{r},t)\cdot\delta\Pi_{kl}(\mathbf{r}',t')\rangle = 2k_{\mathrm{B}}T\eta\left(\delta_{ik}\delta_{jl} + \delta_{il}\delta_{jk}\right)\delta(\mathbf{r}-\mathbf{r}')\delta(t-t'). \tag{4}$$

Here $k_{\mathrm{B}}$ is Boltzmann's constant and $\eta = \rho\nu$ is the dynamic viscosity. The solution of the fluctuating hydrodynamics equations depends on the Reynolds number $\mathrm{Re} = \gamma L^2/\nu$. By adopting the incompressible-flow assumption, Eq. (2), we are neglecting any possible contributions from sound modes.

It is convenient to use dimensionless variables with spatial coordinates $\mathbf{r}(x,y,z)$ in terms of $L$, $t$ in terms of $\gamma^{-1}$. **v** in terms of $L\gamma$, and $\Pi$ in terms of $\rho L^2\gamma^2$. As shown in previous publications [14, 15], by applying a single rotational and a double rotational, one eliminates pressure fluctuations and obtains from Eqs. (2) and (3) two dimensionless fluctuating hydrodynamics equations, one for the fluctuations $\delta v_z$ of the wall-normal component of the velocity and one for the fluctuations $\delta\omega_z = \partial_y\delta v_x - \partial_x\delta v_y$ of the wall-normal component of the vorticity:

$$\frac{\partial}{\partial t}(\nabla^2\delta v_z) + z\frac{\partial}{\partial x}(\nabla^2\delta v_z) - \frac{1}{\mathrm{Re}}\nabla^4(\delta v_z) = [\boldsymbol{\nabla}\times\boldsymbol{\nabla}\times\{\boldsymbol{\nabla}\delta\Pi\}]_z, \tag{5}$$

$$\frac{\partial}{\partial t}(\delta\omega_z) + z\frac{\partial}{\partial x}(\delta\omega_z) - \frac{\partial}{\partial y}\delta v_z - \frac{1}{\mathrm{Re}}\nabla^2(\delta\omega_z) = [\boldsymbol{\nabla}\times\{\boldsymbol{\nabla}\delta\Pi\}]_z. \quad (6)$$

Equation (5) is the stochastic version of what is known as the Orr-Sommerfeld equation and Eq. (6) is the stochastic version of what is known as the Squire equation in the fluid mechanics literature [27, 30].

A procedure for solving these fluctuating equations has been developed in some previous publications [14, 15]. The solution of Eqs. (5) and (6) depends on the Reynolds number Re and on the boundary conditions at the two horizonal plates. At a given shear rate $\gamma$ large $L$ corresponds to large Re and small $L$ corresponds to small Re. For large $L$ and, hence, for large Re (assuming that the laminar average flow is still stable) we can obtain an approximate solution by neglecting the boundary conditions. Then the fluctuating Orr-Sommerfeld Eq. (5) and the fluctuating Squire Eq. (6) can be solved by applying a Fourier transform in terms of a 3-dimensional wave vector $\mathbf{q}(q_x, q_y, q_z)$. As shown in Appendix A, relatively simple expressions are obtained for the nonequibrium part of the equal-time correlation functions in momentum space for the fluctuations of the wall-normal velocity, $\langle\delta v_z^*(\mathbf{q})\delta v_z(\mathbf{q}')\rangle_{\mathrm{NE}}$, of the vorticity, $\langle\delta\omega_z^*(\mathbf{q})\delta\omega_z(\mathbf{q}')\rangle_{\mathrm{NE}}$, as well as for the cross-correlation, $\langle\delta v_z^*(\mathbf{q})\delta\omega_z(\mathbf{q}')\rangle_{\mathrm{NE}}$:

$$\langle\delta v_z^*(\mathbf{q})\delta v_z(\mathbf{q}')\rangle_{\mathrm{NE}} = C_{zz}^{\mathrm{NE}}(\mathbf{q})(2\pi)^3\delta(\mathbf{q}-\mathbf{q}'), \quad (7)$$

$$\langle\delta\omega_z^*(\mathbf{q})\delta\omega_z(\mathbf{q}')\rangle_{\mathrm{NE}} = W_{zz}^{\mathrm{NE}}(\mathbf{q})(2\pi)^3\delta(\mathbf{q}-\mathbf{q}'), \quad (8)$$

$$\langle\delta v_z^*(\mathbf{q})\delta\omega_z(\mathbf{q}')\rangle_{\mathrm{NE}} = \mathrm{i}B_{zz}^{\mathrm{NE}}(\mathbf{q})(2\pi)^3\delta(\mathbf{q}-\mathbf{q}'). \quad (9)$$

Disregarding any boundary conditions makes the equal-time correlations translationally invariant in the three spatial directions, so that their Fourier-transforms are proportional to 3-dimensional delta functions $\delta(\mathbf{q}-\mathbf{q}')$. Explicit expressions for the functions $C_{zz}^{\mathrm{NE}}(\mathbf{q})$ and $W_{zz}^{\mathrm{NE}}(\mathbf{q})$ were presented in previous publications [14, 15] and are reproduced by Eqs. (39) and (40) in Appendix A. Following the same procedure we have also obtained the explicit expression of the cross-correlation $B_{zz}^{\mathrm{NE}}(\mathbf{q})$, as represented by Eq. (41) in Appendix A. The corresponding equal-time correlation functions in momentum space for the fluctuations of the stream-wise and span-wise velocity components are then readily obtained by noting that

$$\delta v_x = \frac{-1}{q_\parallel^2}\left(q_x q_z \delta v_z - \mathrm{i}q_y\delta\omega_z\right), \quad (10)$$

$$\delta v_y = \frac{-1}{q_\parallel^2}\left(q_y q_z \delta v_z + \mathrm{i}q_x\delta\omega_z\right), \quad (11)$$

yielding

$$\langle\delta v_x^*(\mathbf{q})\delta v_x(\mathbf{q}')\rangle_{\mathrm{NE}} = C_{xx}^{\mathrm{NE}}(\mathbf{q})(2\pi)^3\delta(\mathbf{q}-\mathbf{q}'), \quad (12)$$

$$\left\langle \delta v_y^*(\mathbf{q})\delta v_y(\mathbf{q}')\right\rangle_{\mathrm{NE}} = C_{yy}^{\mathrm{NE}}(\mathbf{q})(2\pi)^3\delta(\mathbf{q}-\mathbf{q}'). \tag{13}$$

The functions $C_{xx}^{\mathrm{NE}}(\mathbf{q})$ and $C_{yy}^{\mathrm{NE}}(\mathbf{q})$ are directly related to the functions $C_{zz}^{\mathrm{NE}}(\mathbf{q})$, $W_{zz}^{\mathrm{NE}}(\mathbf{q})$, and $B_{zz}^{\mathrm{NE}}(\mathbf{q})$ in Eqs. (7)-(9) as shown in Eqs. (43) and (44) in Appendix A. Finally, the intensity of the velocity fluctuations in real space, to be substituted into the right-hand side of Eq. (1) for the nonequilibrium pressure tensor $\delta\mathrm{P}(\mathbf{r})$, are obtained by integrating the correlation functions, Eqs. (7), (12) and (13), over the wave vector $\mathbf{q}(q_x,q_y,q_z)$. The results thus obtained for the shear-induced elements $\delta p_{ij}$ of $\delta\mathrm{P}(\mathbf{r})$ are presented in Appendix A and will be further discussed in the subsequent section. These contributions to the nonequilibrium pressure tensor in Eq. (1) do not depend explicitly on the position **r**, since we have assumed the temperature and density to be uniform in space.

However, for confined liquid layers with a finite $L$ and, hence, for finite values of Re, it is necessary to account for the boundary conditions at the two horizontal surfaces. As explained in previous publications [14, 15], in that case we can apply a two-dimensional Fourier transform in terms of a two-dimensional wave vector $\boldsymbol{q}_{\parallel}(q_x,q_y)$ parallel to the horizontal walls, while the dependence of the solution on the coordinate in the wall-normal z-directions needs to be treated separately to account for the boundary conditions at $z=\pm L$. Especially for periodic boundary conditions, this can be simply accomplished by taking advantage of the same solutions obtained without boundary conditions, as shown in Appendix B. The idea is to convert the correlation functions to real space by restricting the allowed $q_z$ values to multiples of $N\pi$ (in dimensionless units). Alternatively, this can be understood as applying a finite sine transform in the z-direction [13]. That is, the intensity of the velocity fluctuations in real space are obtained by integrating these correlation functions over the two-dimensional wave vector $\boldsymbol{q}_{\parallel}(q_x,q_y)$ and a summation over the finite sine transform in the z-direction.

The mathematical details for obtaining the solutions in the absence of boundary conditions, thus for large Re, are presented in Appendix A. The solutions including the boundary effects for a finite-size system in the limit of small Re are presented in Appendix B. In Appendix C we present an analysis of the crossover behavior from small to large Re explicitly for the wall-normal component of the nonequilibrium pressure tensor. Most importantly, we find that the inclusion of finite-size effects is essential for all values of Re corresponding to laminar flow conditions, as further discussed below.

## 3 Fluctuation-induced pressures in a liquid under steady shear

As pointed by out by previous investigators [10-13], and confirmed by Eq. (52) in Appendix A, in the absence of boundary conditions the magnitudes of the elements $\delta p_{ii}$ of the nonequilibrium pressure tensor are proportional to $k_{\mathrm{B}}T(\gamma/\nu)^{3/2}$, or to $(\mathrm{Re})^{3/2}$ in dimensionless form. However, when one accounts for finite-size effects by the imposition of boundary conditions, the shear-rate dependence of all elements $\delta p_{ij}$ of the nonequilibrium pressure tensor changes and, in the limit $\mathrm{Re}\to 0$, they become proportional to $(\mathrm{Re})^2$ in dimensionless form, as shown in Eqs. (60) –(62) in Appendix B. Since the solutions of the dimensionless fluctuating Eqs. (5) and (6) only depend on the Reynolds number Re, we conclude that for arbitrary Re the elements of the nonequilibrium pressure tensor for arbitrary Re will be of the form

$$\delta p_{ij} = V_{ij}^{\infty} k_{\mathrm{B}} T \left(\frac{\gamma}{\nu}\right)^{3/2} \varphi_{ij}(\mathrm{Re}), \tag{14}$$

where $\varphi_{ij}(\mathrm{Re})$ defines a crossover function, such that for fixed $\gamma$ and sufficiently large $L$ $\varphi_{ij}(\mathrm{Re})$ approaches unity, while for fixed $\gamma$ and small $L$ $\varphi_{ij}(\mathrm{Re})$ approaches $(V_{ij}^{0}/V_{ij}^{\infty})(\mathrm{Re})^{1/2}$. Specifically, the two limiting cases may be written as

$$\delta p_{ij}^{\infty} \equiv \lim_{\mathrm{Re}\to\infty} \delta p_{ij} = V_{ij}^{\infty} k_{\mathrm{B}} T \left(\frac{\gamma}{\nu}\right)^{3/2}, \tag{15}$$

$$\delta p_{ij}^{0} \equiv \lim_{\mathrm{Re}\to 0} \delta p_{ij} = V_{ij}^{0} k_{\mathrm{B}} T L \left(\frac{\gamma}{\nu}\right)^{2}. \tag{16}$$

In these equations $V_{ij}^{\infty}$ and $V_{ij}^{0}$ are numerical coefficients, which follow from the solutions of the fluctuating hydrodynamics equations evaluated in Appendix A and Appendix B, respectively.

From the solutions of the fluctuating hydrodynamics equations in the absence of boundary conditions evaluated in Appendix A, we find from Eqs. (46) and (47) for the values of the coefficients $V_{ii}^{\infty}$ in Eq. (15):

$$V_{xx}^{\infty} = +0.0847, V_{yy}^{\infty} = +0.0173, V_{zz}^{\infty} = +0.0106. \tag{17}$$

(See also Table I). From the solutions in the presence of periodic boundary conditions, evaluated in Appendix B, we find for from Eqs. (53)-(55) for the coefficients $V_{ii}^{0}$ in Eq. (16):

$$V_{xx}^{0} = +0.001243,\ V_{yy}^{0} = +0.000414,\ V_{zz}^{0} = +0.000553. \tag{18}$$

Upon substituting the results quoted above for $V_{ii}^{\infty}$ and $V_{ii}^{0}$ into Eqs. (15) and (16) we obtain:

$$\delta p^{\infty} = \frac{1}{3}\sum_i \delta p_{ii}^{\infty} = +0.0375 k_{\mathrm{B}} T \left(\frac{\gamma}{\nu}\right)^{3/2}, \tag{19}$$

$$\delta p^{0} = \frac{1}{3}\sum_i \delta p_{ii}^{0} = +0.000737 k_{\mathrm{B}} T L \left(\frac{\gamma}{\nu}\right)^{2}. \tag{20}$$

In addition to the asymptotic expressions, given by Eqs. (15)-(20) above, we have also determined numerically the dependence of the crossover function $\varphi_{zz}(\mathrm{Re})$ of the wall-normal shear-induced pressure component $\delta p_{zz}$ as a function the Reynolds number Re in Appendix C. The crossover function $\varphi_{zz}(\mathrm{Re})$ thus obtained is shown in Fig. 2. The information in this figure shows that, for Re values corresponding to laminar flow (Re<350 [31]), finite-size effects are always very significant and the limiting solution in the absence of boundary conditions, Eq. (15), is actually never reached in stable laminar-flow. For values of Re corresponding to stable laminar flows the asymptotic solution $\varphi_{zz}(\mathrm{Re}) \propto (\mathrm{Re})^{1/2}$ yields a quadratic dependence of $\delta p_{zz}$ on the shear rate $\gamma$ in accordance with Eq. (16) for $\delta p_{zz}^{0}$. This low-Re solution in the presence of boundary conditions appears to be a better approximation than the asymptotic solution $\varphi_{zz}(\mathrm{Re}) = 1$ in the absence of boundary conditions. The most important conclusion is that that finite-size affects are always important and that a dependence of the pressure enhancements from long-range velocity fluctuations on $\gamma^{3/2}$, predicted by previous authors [10-13], will never be seen in practice.

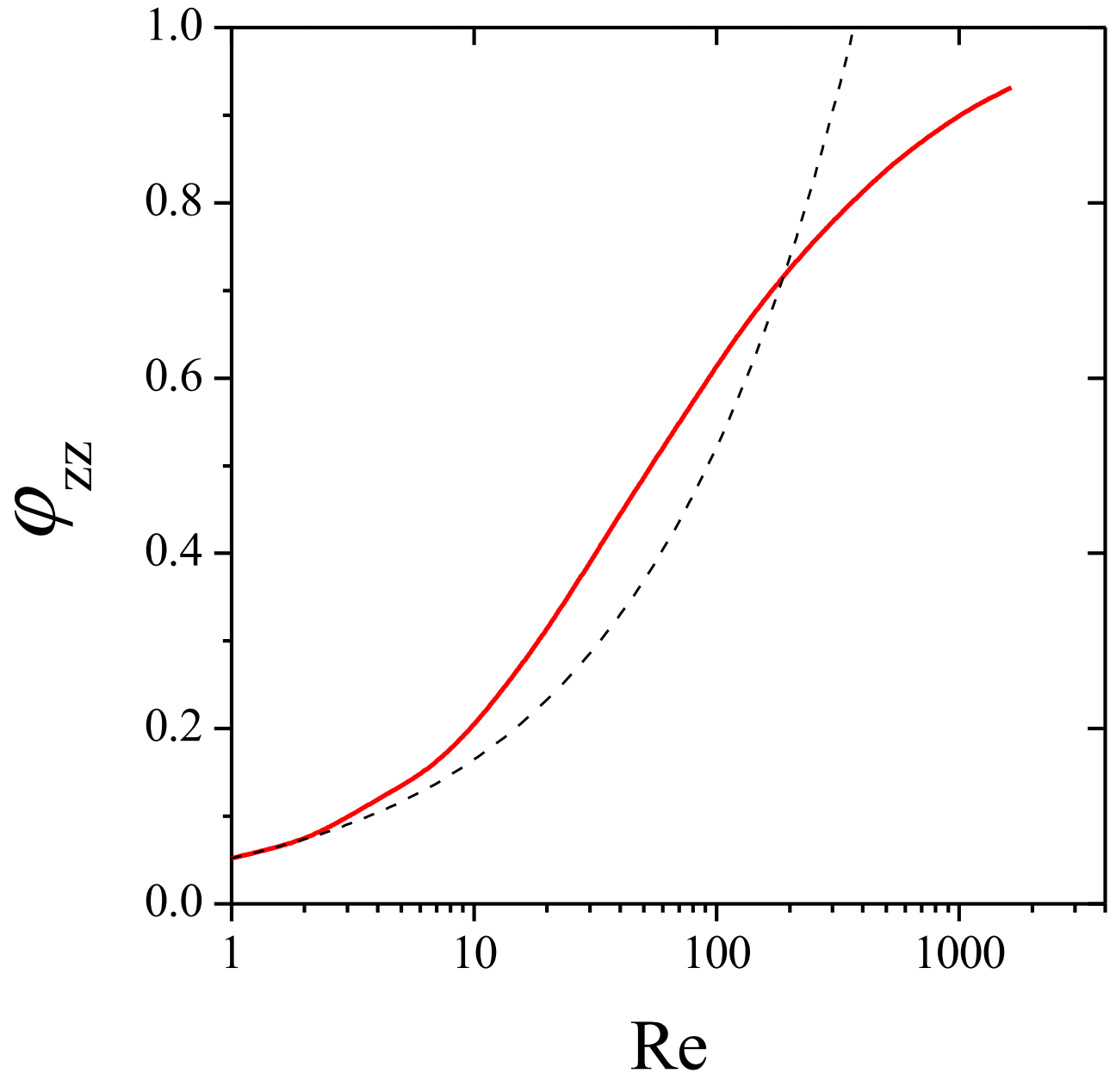

**Fig. 2.** The crossover function $\varphi_{zz}(\mathrm{e})$ for the wall-normal component $\delta p_{zz}$ of the nonequilibrium pressure tensor as a function of the Reynolds number Re. The dashed curve represents the limiting low-Re behavior $\varphi_{zz}(\mathrm{Re}) \propto \mathrm{Re}^{1/2}$ in Eq. (14), yielding the asymptotic Eq. (16) for $\delta p_{zz}$. The limiting large-Re behavior is $\varphi_{zz}(\mathrm{Re}) = 1$, yielding Eq. (15) for $\delta p_{zz}$.

We conclude this section by noting that, while the shear-induced pressures in the limit $L \to \infty$ exhibit a nonanalytic dependence on the shear rate $\gamma$ as $\gamma^{3/2}$, finite-size effects cause a crossover to an analytic dependence of the shear-induced pressures as $\gamma^{2}$.
We also note that coupled sound modes in the limit $U = L\gamma \to \infty$ formally give a result similar to Eq. (15) [10-12]. However, for finite systems considered here the sound-mode contributions are suppressed by a factor $(U/c)^{1/2}$, where *c* is the speed of sound [32], and, hence, we neglect these contributions, having adopted the incompressible-flow assumption, Eq. (2), in formulating the relevant fluctuating hydrodynamics equations

It is interesting that the shear-induced pressure $\delta p^0$ given by Eq. (16) for small Re caused by the finite-size effects *increases* with $L$ at a constant shear rate $\gamma = U/L$, but *decreases* with $L$ at a constant velocity $U$. This is similar to the fluctuation-induced pressure in a liquid subjected to a temperature gradient that increases with $L$ at a constant temperature gradient $\nabla T = \Delta T/L$, but decreases with $L$ at a constant temperature difference $\Delta T$ [26]. However, the difference is that in the case of a temperature gradient the nonequilibrium Casimir pressure is rigorously proportional to $(\nabla T)^2$ and no higher-order gradient terms appear that cause a crossover to a nonanalytic dependence on the gradient [26].

## 4 Interpretation of long-ranged pressure contributions

Attempts to determine the shear-induced pressure tensor in the absence of boundary conditions have been made by Kawasaki and Gunton [10] and by Yamada and Kawasaki [11]. While they did find that the shear-induced pressure varies with the shear rate as $\gamma^{3/2}$, the numerical values of the coefficients are substantially different from the values found by us as shown in Table 1.

**Table 1**. Comparison with literature

| | $V_{xx}^{\infty}$ | $V_{yy}^{\infty}$ | $V_{zz}^{\infty}$ |
|---|---|---|---|
| Kawasaki and Gunton [10] | +0.0050 | –0.0046 | –0.0017 |
| Yamada and Kawasaki [11] | +0.0428 | +0.0173 | +0.0106 |
| This work | +0.0847 | +0.0046 | +0.0106 |

Ernst *et al.* [12] have determined the traceless part of the shear-induced pressure tensor using a kinetic-theory approach. Our results for the traceless part of the shear-induced pressure tensor are in perfect agreement with those obtained by Ernst *et al.* as shown in Table 2. In Appendix A we have also obtained the coefficient $V_{xz}^{\infty} = +0.00916$ for the off-diagonal pressure element $\delta p_{xz}$ in Eq. (15), again in perfect agreement with the result obtained by Ernst *et al*. as shown in Table 2. The magnitude of this fluctuation-induced contribution to the shear viscosity is negligibly small as shown by Ernst *et al*. [12]. It is, therefore, not further discussed in this paper. Wada and Sasa [13] have only determined the wall-normal component of the shear-induced pressure tensor. They find $V_{zz}^{\infty}$ = +0.0106 in the absence of boundary conditions in perfect agreement with our result, but their value $V_{zz}^{0}$ = +0.0002763 for periodic boundary conditions differs from our result exactly by a factor 2.

**Table 2**. Traceless part of shear-induced pressure tensor

| | $V_{xx}^{\infty} - \frac{1}{3}\sum_i V_{ii}^{\infty}$ | $V_{yy}^{\infty} - \frac{1}{3}\sum_i V_{ii}^{\infty}$ | $V_{zz}^{\infty} - \frac{1}{3}\sum_i V_{ii}^{\infty}$ | $V_{xz}^{\infty}$ |
|---|---|---|---|---|
| Ernst *et al.* [12] | +0.0470 | −0.0202 | −0.0268 | +0.00916 |
| This work | +0.0472 | −0.0202 | −0.0269 | +0.00916 |

To estimate the magnitude of the shear-induced pressure enhancement we consider water, which is the liquid commonly used in Couette-flow experiments [31, 33-39]. The smallest gap width thus far employed is about 1.5 mm [33]. The possible experimental plate velocities $U$ may be up to $0.5\ \mathrm{ms}^{-1}$ [40]. A gap width of 1 mm ($L$ = 0.5 mm) and plate velocities $U = \pm 0.5\ \mathrm{ms}^{-1}$ $(\gamma = 1000\ \mathrm{s}^{-1})$ imply Re $\approx$ 280, which is still below the critical Reynolds number for the onset of turbulence [31]. Substituting $\nu = 8.93 \times 10^{-7}\ \mathrm{m^2s^{-1}}$ for the kinematic viscosity of water at 298.15 K [41] into Eqs. (17) and (18) we obtain the estimates

$$\delta p^{\infty} = 6 \times 10^{-9}\mathrm{Pa} \quad \text{and} \quad \delta p^{0} = 2 \times 10^{-9}\mathrm{Pa}, \tag{21}$$

*i.e.,* the shear-induced pressure enhancement is somewhere between $10^{-9}$ and $10^{-8}$ Pa. It is interesting to compare this shear-induced pressure enhancement with those in a liquid layer with the same gap width, either from critical fluctuations $\delta p \cong -2 \times 10^{-11}$ Pa (from Ref. [42], corrected for a sign error) or from nonequilibrum temperature fluctuations caused by the presence of a temperature gradient (25 K /mm) $\delta p \cong 5 \times 10^{-4}$ Pa [26]. We see that the shear-induced pressure enhancement is many orders of magnitude smaller than the Casimir pressures induced by the presence of a temperature gradient. One reason is that temperature fluctuations decay more slowly than velocity fluctuations and, hence, are more strongly impacted by the presence of a temperature gradient. Another reason is that the shear-induced pressure enhancement has a kinetic origin, while the pressure enhancement from a temperature gradient has a potential origin that in liquids is several orders of magnitude larger.

As pointed out above, an important difference between the giant Casimir pressures in liquids subjected to a temperature gradient [26] and the Casimir pressures in the presence of shear, is that the former are orders of magnitudes larger than the shear-induced Casimir pressures given by Eq. (21). Hence, while in the case of a temperature gradient short-range contributions can be neglected, this is no longer obvious in the case of shear-induced pressures enhancements. To estimate a possible contribution from short-range correlations, we note from nonequilibrium statistical mechanics that $\delta p = \kappa\gamma^2$, where $\kappa$ is a nonlinear Burnett coefficient. These nonlinear Burnett coefficients are known to diverge as $L \to \infty$ [43]. Just as in the case of a temperature gradient [26], we may decompose this Burnett coefficient as the sum of a finite short-range contribution $\kappa^{(0)}$ and a long-range contribution $L\kappa^{(1)}$, yielding a short-range (SR) and a long-range (LR) contribution to the shear-induced pressure enhancement:

$$\delta p = \delta p_{\mathrm{SR}} + \delta p_{\mathrm{LR}}, \tag{22}$$

where $\delta p_{\rm SR} = \kappa^{(0)}\gamma^2$ and $\delta p_{\rm LR} = L\kappa^{(1)}\gamma^2$. Comparing with Eq. (20), we note that the shear-induced Casimir pressure, discussed in the previous section, arises from the same long-wavelength hydrodynamic modes that cause the nonlinear Burnett coefficient $\kappa$ to diverge. A complete kinetic theory for the nonlinear Burnett coefficients of real fluids is not available, but it is possible to get an order-of-magnitude estimate for the SR contribution by extending the theory of Enskog for the transport properties of a dense gas of hard spheres to the quadratic level [44]. Starting from an expression for the pressure tensor of a gas of hard spheres provided by Dufty [45] and retaining only the collisional-transfer contribution, which is the dominant one at high densities, we obtain

$$\delta p_{\rm SR} \cong \rho\sigma^2 n\sigma^3 \frac{7\pi}{45}\chi\gamma^2, \tag{23}$$

where $\sigma$ is the hard-sphere diameter, *n* the number density, and $\chi$ the value of the radial distribution function at contact between the spheres. Since for liquid water $\rho = nm = 10^3\ \mathrm{kg\ m^{-3}}$, $m = 3\times10^{-26}\ \mathrm{kg}$, $\sigma = 3\times10^{-10}\ \mathrm{m}$ [41], and estimating $\chi \cong 5$ for a dense liquid, we then conclude from Eq. (16) that for water with $L = 0.5\ \mathrm{mm}$ and $U = 0.5\ \mathrm{ms^{-1}}$ $(\gamma = 1000\ \mathrm{s^{-1}})$ :

$$\delta p_{\rm SR} \cong\ 2\times10^{-10}\ \mathrm{Pa}. \tag{24}$$

On comparing Eq. (24) with Eq. (21) we see that the SR contribution to the induced-pressure enhancement is somewhat smaller than the LR contribution to the shear-induced pressure enhancement, but it is not negligible even at a gap width as large as *L* = 0.5 mm. The SR contribution becomes even more important at smaller values of *L*. From Eq. (23) it follows that, for a fixed velocity *U*, $\delta p_{\rm SR}$ will increase as $L^{-2}$, while $\delta p_{\rm LR}$, due to the long-range velocity fluctuations, will only increase either as $L^{-3/2}$ for large values of Re in accordance with Eq. (19) or even less as $L^{-1}$ for small values of Re in accordance with Eq. (20).

## 5 Computer simulations and nanoscale contributions

A number of computer simulations of model fluids under shear have been reported in the literature [18-24] in an attempt to check a possible dependence of the shear-induced pressure enhancements on $\gamma^{3/2}$ in Eq. (15), predicted in the absence of boundary conditions. Investigators have either claimed to have found agreement [18, 19] or disagreement [20-25] with the prediction of Eq. (15). However, there are two problems with the manner in which these simulation results have been interpreted. The first problem is that the computer simulations probe small nanoscale lengths at extremely large shear-rates $\gamma \approx 10^{11} - 10^{12}\ \mathrm{s^{-1}}$. In addition to the contributions from the short-range fluctuations discussed in the previous section, at these small lengths and high shear rates, there are some other molecular-scale contributions to the calculated shear-induced pressure enhancements. The second problem is that, even in the absence of short-range correlations, in confined liquid layers the dependence on $\gamma^{3/2}$, which is obtained in the absence of boundary conditions, will never be observed under laminar-flow conditions, as was also explained in Section 4.

The first molecular dynamics (MD) simulations on a 3-dimensional sheared fluid consisting of a small number of Lennard-Jones (LJ) particles were performed by Evans [18]. He found results that seemed, especially near the triple point, to indicate a nonequilibrium (NE) pressure enhancement that was proportional to $\gamma^{3/2}$, but with a coefficient that was much larger than the coefficient to be expected from Eq. (17). He noted a similarity with the so-called molasses tail observed in MD simulations of the equilibrium stress-tensor time-correlation function that determines the shear viscosity [46]. It turns out that in this time-dependent correlation function, again near the triple point of LJ particles or near freezing of hard-sphere particles, an apparent long-time tail proportional to $1/t^{3/2}$ appears, but with a coefficient, again, several orders of magnitude larger than the theoretically expected long-time tail coefficient. It was subsequently realized that this so-called molasses tail was not due to long-wave length mode-coupling (MC) effects, but was due to molecular-scale MC effects related to structural relaxation in dense fluids [47-50]. For a review of these molecular-scale MC effects, the reader is referred to a forthcoming book of Dorfman *et al.* [32]. The molecular-scale effects will not only depend on the intermolecular potential adopted, but, at a given density, also on the number of free paths sampled, and, hence, on the number of particles used in the simulations.

All subsequent molecular dynamics simulation studies currently available [19-24] have ignored the effects of molecular-scale correlations that are dominant at nanoscales. Lee and Cumming [19, 20] found an enhancement $\propto \gamma^{3/2}$. But without checking the coefficient, they assumed to have found agreement with both the results of Evans [18] and with Eq. (17), which is impossible as explained above. The more recent MD studies of Sadus and coworkers [21-24] have found effective exponents for the shear-rate dependence ranging from 1.5 to 2 without any theoretical analysis of the results.

The theoretical expression, Eq. (14), for the shear-induced pressure enhancement follows from a solution of the fluctuating hydrodynamics equations for the long-range velocity fluctuations. Numerical solutions of the fluctuating hydrodynamics equations have been obtained some years ago with a direct simulation Monte Carlo method [51, 52] and, more recently, by Varghese *et al.* [25] with a multiparticle collision dynamics method [53-55]. These approaches apply either to dilute gases [51, 52] or to a model fluid with an ideal-gas equation of state [25], but have the merit of evaluating the Casimir pressure purely mechanically, from momentum exchange in particle-wall collisions. And indeed the simulated pressure enhancements found by Varghese *et al*. [25] are of the same order of magnitude estimated from either Eq. (19) or Eq. (20). The calculated pressure enhancement obtained over about one decade of the shear rate seems to scale as $\gamma^2$ and not as $\gamma^{3/2}$. Indeed, in the confined liquid layers considered by Varghese *et al.* the finite-size effects are expected to be very significant and will cause a dependence of the shear-induced pressure enhancements closer to $\gamma^2$ as was elucidated in Section 4. For a quantitative analysis of these type of simulation results we need to determine all three crossover functions $\varphi_{ii}$ $(i = x, y, z)$ in Eq. (14). Such an analysis of the simulation data is outside the scope of the present paper. Moreover, to probe the predicted crossover behavior it would be desirable to pursue these computations over a larger range of gap widths and shear rates.

## 6 Viscous heating

In the derivation of the shear-induced pressure enhancements we have solved the fluctuating-hydrodynamics equation assuming isothermal flow as is commonly done in the statistical mechanics of shear flow [10-13]. That is, possible viscous heating effects have been neglected. This condition is commonly satisfied in computer simulations by special dynamical rules keeping the temperature constant [56] However, this is not a realistic assumption in actual experimental situations.

For shear flow with the velocity gradient in the *z* direction and the fluid velocity $\mathbf{v}_0$ in the *x*-direction, the equation for the rate of change of the temperature is given by [29]

$$\varrho c_p \left[\frac{\partial T}{\partial t} + \mathbf{v}_0 \cdot \nabla T\right] = \lambda \nabla^2 T + \eta \gamma^2 \ , \tag{25}$$

where $c_p$ is the isobaric specific heat capacity and $\lambda$ the thermal conductivity coefficient. In the stationary state $\partial T/\partial t = 0$ and $\mathbf{v}_0 \perp \nabla T$, so that Eq. (25) reduces to

$$\frac{d^2 T}{dz^2} = -\frac{\eta \gamma^2}{\lambda}, \tag{26}$$

as indeed commonly used for non-isothermal plane Couette flow in the literature [57, 58]. Subject to the boundary conditions $T(z = \pm L) = T_0$, the solution of Eq. (26) becomes

$$T(z) - T_0 = +\frac{\eta}{2\lambda}\gamma^2 (L^2 - z^2). \tag{27}$$

Mechanical equilibrium requires

$$\frac{dp}{dz} = \left(\frac{\partial p}{\partial \rho}\right)_T \frac{d\rho}{dz} + \left(\frac{\partial p}{\partial T}\right)_\rho \frac{dT}{dz} = 0, \tag{28}$$

so that

$$\frac{d\rho}{dz} = \left(\frac{\partial \rho}{\partial T}\right)_p \frac{dT}{dz} = -\left(\frac{\partial \rho}{\partial T}\right)_p \frac{\eta \gamma^2}{\lambda} z \ . \tag{29}$$

Integration of Eq. (29) yields

$$\rho(z) = C - \left(\frac{\partial \rho}{\partial T}\right)_p \frac{\eta \gamma^2}{4\lambda} z^2. \tag{30}$$

The integration constant *C* in Eq. (30) is to be determined by satisfying conservation of mass:

$$\frac{1}{2L}\int_{-L}^{+L} dz\, \rho(z) = \rho_0, \tag{31}$$

so that

$$\rho(z) - \rho_0 = \left(\frac{\partial \rho}{\partial T}\right)_p \frac{\eta \gamma^2}{2\lambda}\left(\frac{1}{3}L^2 - z^2\right), \tag{32}$$

where $\varrho_0$ is the density corresponding to $T = T_0$. We note that

$$p - p(\rho_0, T_0) = \left(\frac{\partial p}{\partial T}\right)_\rho (T - T_0) + \left(\frac{\partial p}{\partial \rho}\right)_T (\rho - \rho_0) = \left(\frac{\partial p}{\partial T}\right)_\rho \left[(T - T_0) - \left(\frac{\partial T}{\partial \rho}\right)_p (\rho - \rho_0)\right]. \tag{33}$$

Substitution of Eqs. (27) and (32) into Eq. (33) yields for the resulting pressure enhancement $\Delta p_{\mathrm{VH}} \equiv p - p_0$:

$$\Delta p_{\mathrm{VH}} = \left(\frac{\partial p}{\partial T}\right)_p \frac{\eta \gamma^2}{3\lambda} L^2. \tag{34}$$

To estimate the pressure enhancement caused by viscous heating we consider again the same experimental configuration previously considered in Sections 4 and 5: *L* = 0.5 mm, $U = 0.5\ \mathrm{ms}^{-1}$ ($\gamma = 1000\ \mathrm{s}^{-1}$). Using the known thermophysical properties of liquid water at $25^0\mathrm{C}$ [41] we find from Eq. (34):

$$\Delta p_{\mathrm{VH}} = 70\ \text{Pa}, \tag{35}$$

which is ten orders of magnitude larger than any pressure fluctuation induced enhancement given by Eq. (21). Thus any shear-induced pressure enhancement in experiments will be completely dominated by the effect of viscous heating.

The same problem holds for the pressure enhancements obtained by the currently available computer simulations. As an example we consider here molecular dynamics simulations reported by Lee and Cummings [19, 20] and by Marcelli *et al.* [21] for liquid argon at *T* = 135 K and $\varrho = 1418\ \mathrm{kg\ m}^3$. In terms of dimensional quantities we conclude from Table I and Fig. 2 in the paper of Marcelli *et al.* [21]

$$\delta p = 0.146 \times 10^{-16} (\gamma.\mathrm{s})^2\ \mathrm{Pa}, \tag{36}$$

covering a range of shear rates from $\gamma = 0.04 \times 10^{12}\ \mathrm{s}^{-1}$ till $\gamma = 1.00 \times 10^{12}\ \mathrm{s}^{-1}$. For $\gamma = 0.5 \times 10^{12}\ \mathrm{s}^{-1}$ Eq. (36) implies $\delta p = 3.7 \times 10^6\ \mathrm{Pa}$, in reasonable agreement with the value $\delta p = 3.9 \times 10^6\ \mathrm{Pa}$ from Table II in the paper of Lee and Cummings [20] for the same shear rate. As discussed in Section 4, experimental shear rates are at most $1000\ \mathrm{s}^{-1}$ at which Eq. (36) would suggest a pressure enhancement of less than $10^{-10}\ \mathrm{Pa}$. On the other hand, using the known thermophysical properties of liquid argon [59, 60], we find from Eq. (34) for $\gamma = 1000\ \mathrm{s}^{-1}$ and *L* = 0.5 mm:

$\Delta p_{\rm VH}$=73 Pa, (37)

very similar to the value $\Delta p_{\rm VH}$=70 Pa found in Eq. (35) for liquid water. Thus again, in actual experimental conditions the shear-induced pressure enhancement is completely dominated by the effect of viscous heating.

## 7 Summary

Using nonequilibrium fluctuating hydrodynamics we have demonstrated that the shear-induce nonequilibrium contributions $\delta p_{ij}$ to the pressure tensor resulting from long-range velocity fluctuations satisfy a scaling relation, Eq. (14), in terms of a crossover function that for a given set of boundary conditions depends only on the Re number. For large values of Re, the shear-induced pressure contributions can be obtained from solutions of fluctuating hydrodynamics equations in the absence of boundary conditions. In this limit we have corrected some results for the nonequilibrium pressure enhancement previous reported in the literature by Kawasaki and coworkers [10, 11], while we have found agreement with the value found by Yamada and Kawasaki [11] and by Wada and Sasa [13] for the wall-normal component of the nonequilibrium pressure tensor. For the traceless part of the nonequilibrium pressure tensor we have found complete agreement with the results from kinetic theory previously obtained by Ernst *et al*. [12].

However, we have found that for all values of Re corresponding to actual laminar-flow conditions finite-size effects are very significant and always need to be taken into account. Thus the $\gamma^{3/2}$ dependence on the shear rate $\gamma$ predicted by previous investigators [10-12] will never be observed in practice due to these finite-size corrections.

Molecular dynamics computations, at least in dense fluids, are strongly affected by molecular correlations at nanoscales that have a different physical origin.

Unlike pressure enhancements resulting from either a temperature gradient [26] or a concentration gradient [61], the pressure enhancements caused by velocity fluctuations are very small and negligible in practice.

Finally, we find that, in actual experiments, pressure enhancements resulting from viscous heating are dominant by many orders of magnitude. Hence, while computer simulation of isothermal fluid flow may be useful to check some predictions from statistical physics [25], they are irrelevant for the interpretation of experiments.

## Acknowledgements

We thank J.R. Dorfman for valuable discussions and R. Monchaux for some comments regarding Couette-flow experiments. We are indebted to R.A. Perkins for providing us with the relevant thermophysical-property information for liquid water and liquid argon. The research at the Complutense University was supported by grant ESP2017-83544-C3-2-P of the Spanish *Agencia Estatal de Investigación*. The research at the University of Maryland was supported by the US National Science Foundation under Grant No. DMR-1401449.

## Appendix A: Calculation in the absence of boundary conditions (Re large)

For the calculations it is convenient to use dimensionless variables with position **r** in terms of *L,* wave vector **q** in terms of $L^{-1}$, and velocity $\mathbf{v}$ in terms of $L\gamma$. Then, all the quantities of interest depend only on the Reynolds number and a dimensionless strength of the thermal noise given by [14]:

$$\tilde{S} = \frac{k_B T}{\rho L^2} \frac{1}{\gamma^2 L^2} \frac{1}{\mathrm{Re}}. \tag{38}$$

Large *L* and small *L* at a fixed shear rate $\gamma$ correspond to large $\mathrm{Re}$ and small $\mathrm{Re}$, respectively. As explained in the main text, for large *L* we can neglect the boundary conditions and solve the fluctuating Orr-Sommerfeld and Squire equations by applying a 3-dimensional Fourier transformation. We then obtain for the nonequilibrium (NE) part of the equal-time correlation functions in momentum space Eqs. (7)-(9) with coefficients $C_{zz}^{\mathrm{NE}}(\mathbf{q})$, $W_{zz}^{\mathrm{NE}}(\mathbf{q})$, and $B_{zz}^{\mathrm{NE}}(\mathbf{q})$ that are given by:

$$\frac{C_{zz}^{\mathrm{NE}}(\mathbf{q})}{\tilde{S}\mathrm{Re}} = 2\frac{q_x q_\parallel^2}{q^4} \int_0^\infty d\beta (q_z + q_x\beta)\, e^{-\Gamma(\beta,\mathbf{q})}, \tag{39}$$

$$\frac{W_{zz}^{\mathrm{NE}}(\mathbf{q})}{\tilde{S}\mathrm{Re}} = \frac{q_y^2}{q_x^2} \int_0^\infty d\beta \left[\frac{d\Gamma}{d\beta}\right]^2 [U(\beta,\mathbf{q})]^2\, e^{-\Gamma(\beta,\mathbf{q})}, \tag{40}$$

$$\frac{B_{zz}^{\mathrm{NE}}(\mathbf{q})}{\tilde{S}\mathrm{Re}} = \frac{q_\parallel q_y}{q^2 q_x} \int_0^\infty d\beta \left[\frac{d\Gamma}{d\beta}\right]^2 U(\beta,\mathbf{q})\, e^{-\Gamma(\beta,\mathbf{q})}. \tag{41}$$

with

$$\begin{aligned} \Gamma(\beta,\mathbf{q}) &= \frac{2\beta}{3\mathrm{Re}}(q_x^2\beta^2 + 3\beta q_x q_z + 3q^2), \\ U(\beta,\mathbf{q}) &= \frac{\mathrm{Re}}{2}\left[\operatorname{atan}\left(\frac{q_z + \beta q_x}{q_\parallel}\right) - \operatorname{atan}\left(\frac{q_z}{q_\parallel}\right)\right], \\ &= q_x q_\parallel \int_0^\beta \left[\frac{d\Gamma(u)}{du}\right]^{-1} du \end{aligned} \tag{42}$$

as given by Eq. (39) and Eq. (43b) in Ref. [15], which are exactly the same as Eqs. (39) and (40) above, while the new Eq. (41) for the cross-correlation has been obtained following the same techniques. In these equations $q_\parallel$ is the magnitude of the component $\mathbf{q}_\parallel$ of the wave vector in the *x-y* plane, *i.e.,* parallel to the plates

As explained in the main text, from Eqs. (39)-(41) one can readily obtain also the correlation functions in momentum space for the stream-wise and span-wise components of the velocity fluctuations yielding Eqs. (12) and (13) with coefficients $C_{xx}^{\mathrm{NE}}(\mathbf{q})$ and $C_{yy}^{\mathrm{NE}}(\mathbf{q})$, that are related to the coefficients $C_{zz}^{\mathrm{NE}}(\mathbf{q})$, $W_{zz}^{\mathrm{NE}}(\mathbf{q})$, and $B_{zz}^{\mathrm{NE}}(\mathbf{q})$ by

$$C_{xx}^{\mathrm{NE}}(\mathbf{q}) = \frac{q_x^2 q_z^2}{q_\parallel^4} C_{zz}^{\mathrm{NE}}(\mathbf{q}) + \frac{q_y^2}{q_\parallel^4} W_{zz}^{\mathrm{NE}}(\mathbf{q}) + 2\frac{q_x q_y q_z}{q_\parallel^4} B_{zz}^{\mathrm{NE}}(\mathbf{q}), \tag{43}$$

$$C_{yy}^{\mathrm{NE}}(\mathbf{q}) = \frac{q_y^2 q_z^2}{q_\parallel^4} C_{zz}^{\mathrm{NE}}(\mathbf{q}) + \frac{q_x^2}{q_\parallel^4} W_{zz}^{\mathrm{NE}}(\mathbf{q}) - 2\frac{q_x q_y q_z}{q_\parallel^4} B_{zz}^{\mathrm{NE}}(\mathbf{q}). \tag{44}$$

As discussed in Sect. 2, integration of Eqs. (39), (43), and (44) for $\mathbf{q}\epsilon\mathbb{R}^3$ yields the diagonal elements of $\langle \delta\mathbf{v}\delta\mathbf{v}\rangle_{\mathrm{NE}}$ in real space and for large Re.

As an example, we consider here the computation of $V_{zz}^{\infty}$. In terms of dimensionless units:

$$\tilde{S}\mathrm{Re}V_{zz}^{\infty}(\mathrm{Re})^{\frac{3}{2}} = \frac{1}{(2\pi)^3}\int_{\mathbb{R}^3} C_{zz}^{\mathrm{NE}}(\mathbf{q})d\mathbf{q}. \tag{45}$$

To evaluate the coefficient $V_{zz}^{\infty}$, after substitution of Eq. (39) into Eq. (45), we adopt spherical coordinates for the integration over $\mathbf{q}$. We first integrate over the magnitude $q$ of the vector $\mathbf{q}$, which can be done analytically and yields the prefactor $(\mathrm{Re})^{3/2}$. A second integration over the polar angle can also be performed analytically taking advantage of the symmetry properties of the integral. The final double integral, over the azimuthal angle and over the parameter $\beta$, can be simplified but not performed analytically and has been evaluated numerically:

$$\begin{aligned} V_{zz}^{\infty} &= \frac{\sqrt{3}}{32\pi^3}\,\Gamma\left(\frac{1}{4}\right)^2 \int_0^\infty \frac{d\beta}{\beta^{\frac{3}{2}}}\int_0^\pi \frac{(\beta+\cos\theta)(\sin\theta)^{\frac{9}{2}}}{(\beta^2+3\beta\cos\theta+3)^{\frac{3}{2}}}d\theta, \\ &\simeq 0.0106, \end{aligned} \tag{46}$$

which is the value quoted in Table 1 of the main text. The other coefficients, $V_{xx}^{\infty}$ and $V_{yy}^{\infty}$, have been evaluated in a similar fashion from Eqs. (43) and (44). The resulting values are

$$V_{xx}^{\infty} = +0.0847, V_{yy}^{\infty} = +0.0173, \tag{47}$$

also shown in Table 1, where a detailed discussion and comparison with the literature is presented.

## Appendix B: Calculation for periodic boundary conditions (Re small)

Strictly speaking, the main result of Appendix A, namely, that the intensity of the velocity fluctuations (and associated pressure) is proportional to $(\gamma)^{3/2}$, only applies for spatial points that are very far from the boundaries. In practice, since non-equilibrium fluctuations have a long spatial range, their intensity will be strongly affected by the boundary conditions [6]. To illustrate how this intensity changes due to confinement, we consider in this Appendix periodic boundary conditions (PBC) that are commonly adopted in computer simulations.

According to Eq. (45), the computation of the intensity of the velocity fluctuations in real space without boundary conditions involves an integration over wave vectors $\mathbf{q}\epsilon\mathbb{R}^3$, in particular over the wall-normal *z*-component $q_z \in (-\infty, \infty)$. Following previous authors [13], we perform a calculation for PBC in the *z*-direction by allowing the $q_z$ to take only values which are multiples of $N\pi$ (in dimensionless units), *i.e.*, $q_z = N\pi$ with $N = \pm 1, \pm 2, \pm 3, \ldots$ Alternatively, this approach can be understood as approximating the integral over $q_z$ as a series. For instance, in the case of $C_{zz}^{\mathrm{NE}}(\mathbf{q})$, one approximates the $q_z$ integral on the right-hand side of Eq. (45) as:

$$\frac{1}{(2\pi)}\int_{-\infty}^{\infty} C_{zz}^{\mathrm{NE}}(\mathbf{q}_{||}, q_z)\mathrm{d}q_z \simeq \frac{1}{(2\pi)}\left[\frac{\pi}{L}\sum_{N=1}^{\infty}\left[C_{zz}^{\mathrm{NE}}\left(\mathbf{q}_{||}, \frac{N\pi}{L}\right) + C_{zz}^{\mathrm{NE}}\left(\mathbf{q}_{||}, -\frac{N\pi}{L}\right)\right]\right]. \quad (48)$$

Then, the calculation without boundary conditions of Appendix A corresponds to taking the limit $L \to \infty$ in Eq. (48), while the calculation for PBC corresponds to taking $L = 1$ in Eq. (55), thus forcing $C_{zz}^{\mathrm{NE}}(\mathbf{q}_{||}, q_z)$ to be, in real space, periodic in the *z*-direction. Note that, to obtain the actual intensity of fluctuations, Eq. (48) still needs to be integrated over $\mathbf{q}_{||}\epsilon\mathbb{R}^2$.

Upon substituting Eq. (39) into Eq. (48), and either Eq. (43) or Eq. (44) in the corresponding expressions for $C_{xx}^{\mathrm{NE}}$ and $C_{yy}^{\mathrm{NE}}$, the intensities of the velocity fluctuations can be evaluated for arbitrary $\mathrm{Re}$, and the crossover functions $\varphi_{ii}(\mathrm{Re})$ of Eq. (2) computed. The resulting series and integrals cannot be evaluated analytically in general, but can be studied numerically. As an example of these calculations we explicitly consider again the case of $[C_{zz}^{\mathrm{NE}}]_{\mathrm{PBC}}$. Use of Eq. (48) (with $L = 1$) for substituting the $q_z$ integral on the right-hand side of Eq. (45), after performing analytically the integral over $q_y$ and changing variables in the remaining $q_x$ and $\beta$ integrals, results in:

$$\left[\frac{C_{zz}^{\mathrm{NE}}(\mathrm{Re})}{\tilde{S}\mathrm{Re}}\right]_{\mathrm{PBC}} = \frac{2\mathrm{Re}}{8\pi}\sum_{N=1}^{\infty} F\left(\frac{N^2\pi^2}{\mathrm{Re}}\right), \quad (49)$$

with the function

$$F(u) = \int_0^\infty d\beta \int_{-\infty}^{\infty} dq_x \exp\left[-\frac{2\beta^2 q_x}{3u}\left(\frac{q_x\beta}{u} + 3\right)\right]\frac{1 - \mathrm{erf}\left(\sqrt{1+q_x^2}\sqrt{2\beta}\right)}{\sqrt{1+q_x^2}} \times\left\{2\left(1 + \frac{q_x\beta}{u}\right)\left[q_x + \frac{\beta}{3}\left(\frac{2\beta^2 q_x}{u^2} + 3\frac{\beta}{u} + 6q_x\right)\right] - \frac{\beta}{u}\right\}, \quad (50)$$

where the integrals in Eq. (50) are perfectly converging for any $u \neq 0$ and, correspondingly, the function $F(u)$ is analytic. Notice that the summand in Eq. (49) is the same for $\pm N$. In addition, because of the structure on the right-hand side of Eq. (49), the large $\mathrm{Re}$ limit results in:

$$\left[\frac{C_{zz}^{\mathrm{NE}}(\mathrm{Re})}{\tilde{S}\mathrm{Re}}\right]_{\mathrm{PBC}} \xrightarrow{\mathrm{Re}\to\infty} \frac{(\mathrm{Re})^{\frac{3}{2}}}{8\pi}\int_{-\infty}^{\infty} F(u^2)du \simeq 0.0106\,(\mathrm{Re})^{3/2}, \tag{51}$$

which is equivalent to having performed the full integral over $q_z$ of the original $C_{zz}^{\mathrm{NE}}(\mathbf{q}_{||}, q_z)$ of Eq. (48) and, thus, reproduces Eq. (46) of Appendix A. As mentioned in the main text, the calculation without boundary conditions is equivalent to the calculation for PBC in the limit of very large $\mathrm{Re}$.

From Eqs. (49) and (50) it is clear that, for arbitrary $\mathrm{Re}$ only a numerical calculation is possible. However, in the limit $\mathrm{Re} \to 0$ some analytical progress is feasible. Since

$$F(u) \xrightarrow{u\to\infty} \frac{1}{24u} + O\left(\frac{1}{u^2}\right), \tag{52}$$

one readily obtains

$$\left[\frac{C_{zz}^{\mathrm{NE}}}{\tilde{S}\mathrm{Re}}\right]_{\mathrm{PBC}} \xrightarrow{\mathrm{Re}\to 0} \frac{\mathrm{Re}^2}{576\pi} + O(\mathrm{Re}^4). \tag{53}$$

The other two components of the main diagonal velocity correlations in real space, $[C_{xx}^{\mathrm{NE}}]_{\mathrm{PBC}}$ and $[C_{yy}^{\mathrm{NE}}]_{\mathrm{PBC}}$, can be treated in a similar way. For large $\mathrm{Re}$ we recover the results of Appendix A, while a power series expansion for $\mathrm{Re} \to 0$ yields:

$$\left[\frac{C_{xx}^{\mathrm{NE}}}{\tilde{S}\mathrm{Re}}\right]_{\mathrm{PBC}} \xrightarrow{\mathrm{Re}\to 0} \frac{\mathrm{Re}^2}{256\pi} + O(\mathrm{Re}^4) \tag{54}$$

$$\left[\frac{C_{yy}^{\mathrm{NE}}}{\tilde{S}\mathrm{Re}}\right]_{\mathrm{PBC}} \xrightarrow{\mathrm{Re}\to 0} \frac{\mathrm{Re}^2}{768\pi} + O(\mathrm{Re}^4). \tag{55}$$

These small $\mathrm{Re}$ power series expansions can be alternatively obtained by simply changing the integration variable to $\beta' = \beta/\mathrm{Re}$ in the original Eqs. (39)-(41), and expanding the resulting integrand in powers of $\mathrm{Re}$. As discussed in Section 3, the condition $\mathrm{Re} \to 0$ corresponds to small system size and is more appropriate for the interpretation of computer simulations [55]. The numbers multiplying the $\mathrm{Re}^2$ terms in Eqs. (53)-(55) yield, for PBC, the $V_{ii}^0$ coefficients quoted in Eq. (18), while the value quoted in Eq. (20) is

$$\frac{1}{3}\left[\frac{1}{256\pi} + \frac{1}{768\pi} + \frac{1}{576\pi}\right] \cong 0.000737. \tag{63}$$

A value of $V_{zz}^0 = 1/1152\pi$ has been reported by Wada and Sasa [13] which differs from our result, Eq. (53), exactly by a factor of 2. This difference may be related to the fact that, congruent with the fluid-mechanics literature, we took the size of our layer as $2L$. However, Wada and Sasa [13] have provided only little details about their calculation for PBC, so that an ultimate explanation is not available. Equations (54) and (55) for the other two $V_{ii}^0$ coefficients are presented here for the first time.

It is interesting to note that, of the three terms contributing to $C_{xx}^{\mathrm{NE}}(\mathbf{q})$ in Eq. (43) or to $C_{yy}^{\mathrm{NE}}(\mathbf{q})$ in Eq. (44), the dominant contribution at $\mathrm{Re} \to 0$ comes from the one associated with vorticity fluctuations. And indeed, the vorticity term is the only one contributing to either $V_{xx}^0$ or $V_{yy}^0$. The proportionality to $\mathrm{Re}^2$ at small Re is also obtained when rigid boundary conditions are adopted for the velocity [14, 15], in which case an exact solution is not readily feasible [16, 17] and in practice we have settled for only a low-order Galerkin approximation [14 15]. Here, for PBC, the coefficients multiplying the $\mathrm{Re}^2$ terms in Eqs. (53)-(55) are exact.

## Appendix C: Crossover between small and large Re behavior

We have also evaluated numerically $[C_{zz}^{\mathrm{NE}}]_{\mathrm{PBC}}$from Eq. (49) for all $\mathrm{Re}$ numbers between 1 and 1600 and the corresponding crossover function $\varphi_{zz}$ introduced in Eq. (2). The results are presented in Fig. 3, where the upper panel shows $[C_{zz}^{\mathrm{NE}}]_{\mathrm{PBC}}$ and the lower panel $\varphi_{zz}$ as a function of Re. The convergence of the series (49) is very slow, particularly for large $\mathrm{Re}$, and up to $N = 60$ terms have been added to obtain the values presented in Fig. 3. Added as thin lines in the upper panel are the two asymptotic limits, large and small $\mathrm{Re}$, as given by Eqs. (51) and (53), respectively. We see in the upper panel how $[C_{zz}^{\mathrm{NE}}]_{\mathrm{PBC}}$ crosses over continuously and smoothly between the two analytic asymptotic limits. The asymptotic limit for large $\mathrm{Re}$ (*i.e.*, $\varphi_{zz} = 1$) is indicated in the lower panel.

We emphasize that all results in these appendixes have been obtained from fluctuating hydrodynamics equations *linearized* in the fluctuating velocity: the so-called stochastic Orr-Sommerfeld and Squire equations specified in Section 2. The nonlinear part of the advection term, responsible for transition to turbulence, has been neglected from the outset. Hence, our results only pertain to Reynolds numbers (Re<350) for which plane Couette flow is stable [31]. From the information in the lower panel of Fig. 3 we see that for all Reynold numbers corresponding to laminar flows the crossover function $\varphi_{zz}$ is substantially less than unity. Hence the prediction that $\delta p$ should vary as $\gamma^{3/2}$ in the absence of boundary conditions will never be observed in practice. Taking into account finite-size corrections will be essential in the analysis of data for the shear-induced pressure enhancement in laminar flow.

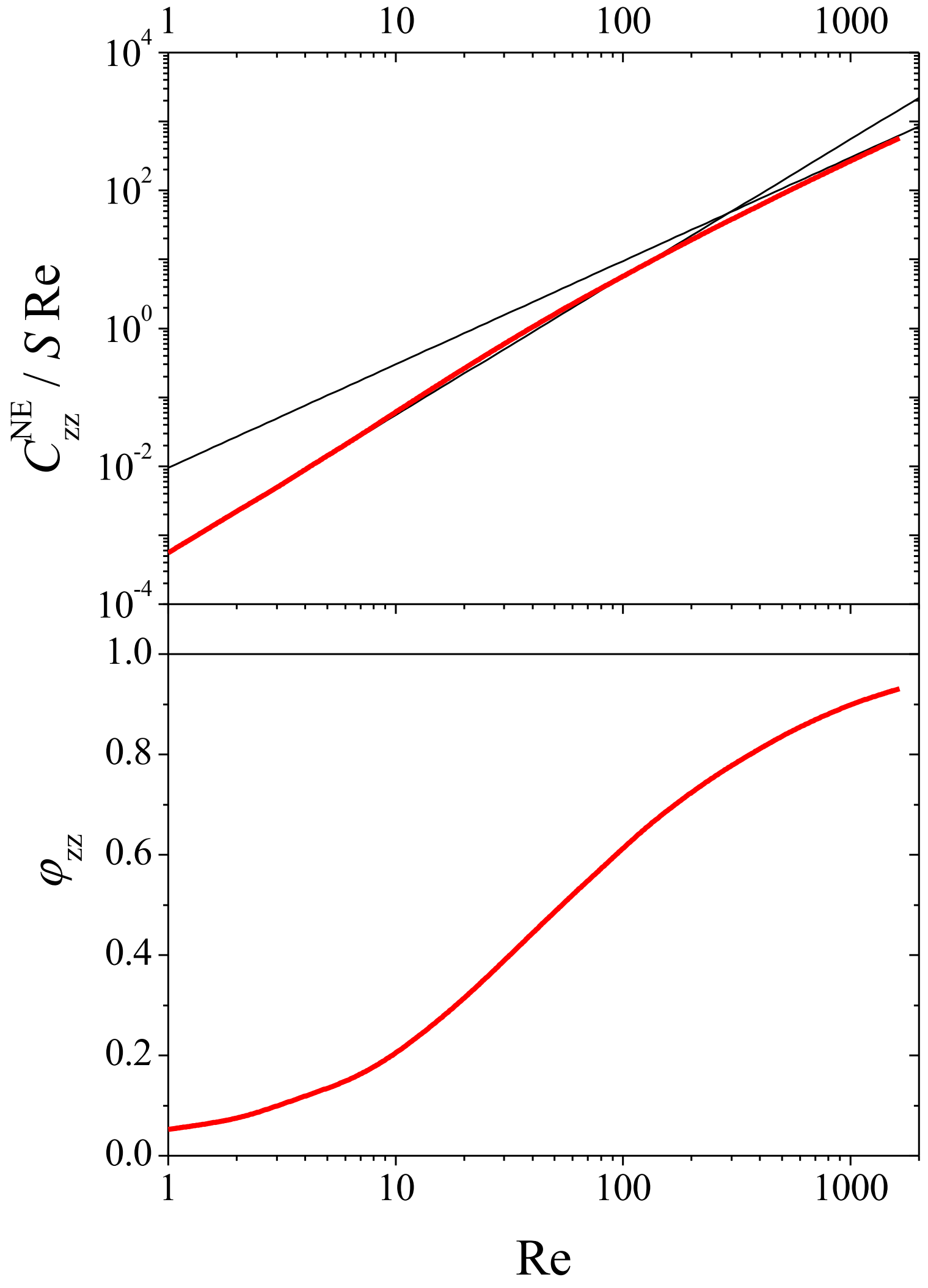

**Fig. 3**: The red curves represent, as a function of the Reynolds number, the intensity of wall-normal velocity fluctuations, $C_{zz}^{\mathrm{NE}}$ (upper panel) and the corresponding crossover function, $\varphi_{zz}$ (lower panel) evaluated numerically from Eqs. (49) and (50). The two thin lines in the upper panel indicate asymptotic limits at small $\mathrm{Re}$, Eq. (45), and at large $\mathrm{Re}$, Eq. (51). In the lower panel the asymptotic limit for large $\mathrm{Re}$ (*i.e.*, $\varphi_{zz} = 1$) is indicated.

## References

1. M. Kardar and R. Golestanian, Rev. Mod. Phys. **71**, 1233 (1999).
2. G. L. Klimchitskaya, U. Mohideen, and V. M. Mostepanenko, Rev. Mod. Phys. **81**, 1827 (2009).
3. M. Krech, *The Casimir Effect in Critical Systems* (World Scientific, Singapore, 1994).
4. T.R. Kirkpatrick, J.K.Bhattacharje, and J.V. Sengers, Phys. Rev. Lett.**119**, 030603 (2017).
5. J. R. Dorfman, T. R. Kirkpatrick, and J. V. Sengers, Annu. Rev. Phys. Chem. **45**, 213 (1994).
6. J. M. Ortiz de Zárate and J. V. Sengers, *Hydrodynamic Fluctuations in Fluids and Fluid Mixtures* (Elsevier, Amsterdam, 2006).
7. J. F. Lutsko and J. W. Dufty, Phys. Rev. A **32**, 3040 (1985).
8. J. F. Lutsko and J. W. Dufty, Phys. Rev. E **66**, 041206 (2002).
9. J. V. Sengers and J. M. Ortiz de Zárate, J. Non-Newtonian Fluid Mech. **165**, 925 (2010).
10. K. Kawasaki and J. D. Gunton, Phys. Rev. A **8**, 2048 (1973).
11. T. Yamada and K. Kawasaki, Progr. Theor. Phys. (Japan) **53**, 111 (1975).
12. M. H. Ernst, B. Cichocki, J. R. Dorfman, J. Sharma, and H. van Beijeren, J. Stat, Phys. **18**, 237 (1978).
13. H. Wada and S. I. Sasa, Phys. Rev. E **67**, 065302(R) (2003).
14. J. M. Ortiz de Zárate and J. V. Sengers, Phys. Rev. E **77**, 026306 (2008).
15. J. M. Ortiz de Zárate and J. V. Sengers, Phys. Rev. E **79**, 046308 (2009).
16. J. M. Ortiz de Zárate and J. V. Sengers, J. Stat. Phys. **144**, 774 (2011).
17. J. M. Ortiz de Zárate and J. V. Sengers, J. Stat. Phys. **150**, 540 (2013).
18. D. J. Evans, Phys. Rev. A **23**, 1988 (1981).
19. S. H. Lee and P. T. Cummings, J. Chem. Phys. **99**, 3919 (1993).
20. S. H. Lee and P. T. Cummings, J. Chem. Phys. **101**, 6206 (1994).
21. G. Marcelli, B. D. Todd, and R. J. Sadus, Phys. Rev. E **63**, 021204 (2001).
22. J. Ge, G. Marcelli, B. D. Todd, and R. J. Sadus, Phys. Rev. E **64**, 021201 (2001).
23. J. Ge, B. D. Todd, G. Wu, and R. J. Sadus, Phys. Rev. E **67**, 061201 (2003).
24. A. Ahmed, P. Mausbach, and R. J. Sadus, Phys. Rev. E **82**, 011201 (2010).
25. A. Varghese, G. Gompper, and R. G. Winkler, Phys. Rev. E **96**, 062617 (2017).
26. T. R. Kirkpatrick, J. M. Ortiz de Zárate, and J. V. Sengers, Phys. Rev. Lett. **110**, 235902 (2013).
27. P.G. Drazin and W.H. Reid, *Hydrodynamic Stability*, 2nd ed. (Cambridge University Press, Cambridge, U.K., 2004).
28. B. Eckhardt and R. Pandit, Eur. Phys. J. B **33**, 373 (2003).
29. L.D. Landau and E.M. Lifshitz, *Fluid Mechanics*, 2nd ed. (Pergamon, London, 1987).
30. P.J. Schmid and D.S. Henningson, *Stability and Transition in Shear Flows* (Springer, Berlin, 2001).
31. N. Tillmark and P. Alfredson, J. Fluid Mech. **235**, 89 (1992).
32. J. R. Dorfman, T.R.Kirkpatrick, and H. van Beijeren, *Contemporary Kinetic Theory of Matter* (Cambridge University Press, Cambridge, to be published).
33. F. Daviaud, J. Hegseth, and P. Bergé, Phys. Rev. Lett. **69**, 2511 (1992).

34. O. Dauchot and E. Daviaud, Phys. Fluids **7**, 335 (1995).
35. S. Bottin, F. Daviaud, P. Manneville, and O. Dauchot, Europhys. Lett. **43**, 171 (1998).
36. A. Prigent, G. Grégoire, H. Chaté, and O. Dauchot, Physica D **174**, 100 (2003).
37. M. Couliou and R. Monchaux, Phys. Fluids **27**, 034101 (2015).
38. L. Klotz, G. Lemoult, I. Frontczak, L. S. Tuckerman, and J. E. Wesfreid, Phys. Rev. Fluids **2**, 043904 (2017).
39. L. Klotz and J. Wesfreid, J. Fluid Mech. **829**, R4 (2017).
40. R. Monchaux, private communication.
41. Revised Supplementary Release on Properties of Liquid Water at 0.1 MPa, IAPWS SR6-08 (2011), available at www.iapws.org.
42. T. R. Kirkpatrick, J. M. Ortiz de Zárate, and J. V. Sengers, Phys. Rev. E **93**, 012148 (2016).
43. J. J. Brey, J. Chem. Phys. **79**, 4585 (1983).
44. H. van Beijeren and J. R. Dorfman, Physica **68**, 437 (1973).
45. J. W. Dufty, Mol. Phys. **100**, 2331 (2002).
46. J. J. Erpenbeck and W. Wood, J. Stat. Phys. **24**, 455 (1981).
47. H. van Beijeren, Phys. Lett. A **105**, 191 (1984).
48. T. R. Kirkpatrick, Phys. Rev. Lett. **53**, 1735 (1984).
49. T. R. Kirkpatrick, Phys. Rev. A **32**, 3130 (1985).
50. T. R. Kirkpatrick and J. C. Nieuwoudt, Phys. Rev. A **33**, 2651 (1986).
51. A. L. Garcia, M. Malek Mansour, G. C. Lie, M. Mareschal, and E. Clementi, Phys. Rev. A **36**, 4348 (1987).
52. F. J. Alexander, A. L. Garcia, and B. J. Adler, Phys. Fluids **6**, 3854 (1994).
53. G. Gompper, T. Ihle, D.M. Kroll, and R.C. Winkler, Adv. Polymer Science 221, 1-87 (2009).
54. C.-C. Huang, A. Varghese, G. Gompper, and R.C. Winkler, Phys. Rev. E **91**, 013310 (2015).
55. A. Varghese, C.-C. Huang, R.G. Winkler, and G. Gompper, Phys. Rev. E **92**, 053002 (2015).
56. J. W. Dufty and J. Lutsko, in *Recent Developments in Nonequilibrium Thermodynamics: Fluids and Related Topics*, Vol. 253 of *Lecture Notes in Physics*, edited by J. Casas-Vázquez, D. Jou, and J. M. Rubí (Springer, Berlin, 1986), pp. 47–84.
57. R.A. Burton, *Heat, Bearings, and Lubrication*, Sproinger, New York, 2000), Ch. 2.
58. R.B.Bird, W.E. Stewart, and E.N. Lightfoot, *Transport Phenomena*, 2nd ed. (Wiley, New York, 2002), Ch. 10.
59. E.W. Lemon, E.W. Bell, M.L. Huber, and M.O. McLinden, *NIST Standard Reference Database 23: Reference Fluid Thermodynamic and Transport Properties-REFPROP, Version 10.0* (National Institute of Standards and Technology, Gaithersburg, MD, 2018).
60. Ch. Teagler, R. Span, and W.A. Wagner, J. Phys. Chem. Ref. Data **28**, 779 (1999).
61. T. R. Kirkpatrick, J. M. Ortiz de Zárate, and J. V. Sengers, Phys. Rev. Lett. **115**, 035901 (2015).